\documentclass[tinyml]{acmart}
\usepackage{dblfloatfix}
\usepackage{comment}
\usepackage{hyperref}

\newcommand{\Asmer}[1]{\textcolor{black}{#1}}
\newcommand{\Mozhgan}[1]{\textcolor{black}{#1}}

\AtBeginDocument{%
  \providecommand\BibTeX{{%
    \normalfont B\kern-0.5em{\scshape i\kern-0.25em b}\kern-0.8em\TeX}}}

\setcopyright{none}
\copyrightyear{2024}
\acmYear{2024}

\begin{document}

\title{Energy-Aware FPGA Implementation of Spiking Neural Network with LIF Neurons }


\author{Asmer Hamid Ali}
\email{aali56@jhu.edu}
\affiliation{%
 \institution{Johns Hopkins University}
 \streetaddress{}
 \city{Baltimore}
 \state{Maryland}
 \country{USA}}

\author{Mozhgan Navardi}
\email{mnavard1@jhu.edu}
\affiliation{%
 \institution{Johns Hopkins University}
 \streetaddress{}
 \city{Baltimore}
 \state{Maryland}
 \country{USA}}

\author{Tinoosh Mohsenin}
\email{tinoosh@jhu.edu}
\affiliation{%
 \institution{Johns Hopkins University}
 \streetaddress{}
 \city{Baltimore}
 \state{Maryland}
 \country{USA}}

\begin{abstract}




\Asmer{Tiny Machine Learning~(TinyML) has become a growing field in on-device processing for Internet of Things~(IoT) applications, capitalizing on AI algorithms that are optimized for their low complexity and energy efficiency. These algorithms are designed to minimize power and memory footprints, making them ideal for the constraints of IoT devices. Within this domain, Spiking Neural Networks~(SNNs) stand out as a cutting-edge solution for TinyML, owning to their event-driven processing paradigm which offers an efficient method of handling dataflow. This paper presents a novel SNN architecture based on the $1^{st}$ Order Leaky Integrate-and-Fire~(LIF) neuron model to efficiently deploy vision-based ML algorithms on TinyML systems.}
\Asmer{A hardware-friendly LIF design is also proposed, and implemented on a Xilinx Artix-7 FPGA. To evaluate the proposed model, a collision avoidance dataset is considered as a case study. The proposed SNN model is compared to the state-of-the-art works and Binarized Convolutional Neural Network~(BCNN) as a baseline. The results show the proposed approach is 86\% more energy efficient than the baseline.}

\end{abstract}



\maketitle

\section{Introduction}
\label{sec:intro}

  

\Mozhgan{Deep learning algorithms based on Artificial Neural Networks~(ANN) are popular in Internet of Things~(IoT) applications due to their AI capabilities. However, the size and complexity of these models require high-performance computing~(HPC) clusters in data centers, which poses challenges for IoT devices that gather data on the edge~\cite{navardi2023mlae2, humes2023squeezed}. Edge computing, where data is processed on the device itself, can address these challenges but is limited by memory and energy constraints~\cite{navardi2023mlae2, tinyvqa, mazumder2023reg, navardi2023metae2rl}. It led Tiny Machine Learning~(TinyML) to emerge which considers efficient deployment of ANN models on tiny robots such as Unmanned Aerial Vehicles~(UAVs) and Unmanned Ground Vehicles~(UGVs)~\cite{manjunath2023reprohrl}. However, deploying ANN models on the edge is still limited by power and hardware constraints due to the highly intensive computational of such models. To overcome this, there is interest in bio-inspired approaches like Spiking Neural Networks~(SNNs), which have lower energy consumption~\cite{nguyen2022hardware,davidson2021comparison, vestias2019survey,roy2017programmable}.}

Recent advancements in the field of neural networks have led to an era where the computational efficiency of SNNs is being harnessed in more innovative and practical ways \cite{introme1,introme2,introme3}. SNNs, with their ability to mimic the intricacies of biological neural networks, represent a significant leap from traditional artificial neural networks. They offer a more nuanced approach to information processing, mimicking the dynamic, temporal characteristics of biological neuron activity.

The integration of SNNs with Field-Programmable Gate Arrays~(FPGAs) has opened new avenues in computing, merging the adaptability of neural networks with the energy efficiency and customization offered by FPGAs \cite{introme4,introme5,introme6,introme7,FPGA_use1,FPGA_use2, mazumder2021survey}. This synergy is crucial, especially in applications where power efficiency is important, such as in embedded systems or portable devices. While recent research has made strides in developing effective FPGA architectures for SNNs and creating more biologically plausible models \cite{Ref4,SNNcompare1, SNNcompare2,SNN}, there remains a gap in applying these advancements to complex, real-world scenarios. Much of the existing work \cite{introme8,introme9,introme10} focuses on pattern recognition tasks using simplified datasets like MNIST \cite{MNIST}, has not fully explored the potential of SNNs in more challenging and realistic environments. This research aims to bridge this gap by shifting the focus to a more demanding and realistic dataset: collision avoidance. The study explores the training of SNNs for collision avoidance and their implementation on FPGA platforms. By moving beyond the conventional realm of simple image classification tasks, this work aims to validate the practicality and robustness of SNNs in high-stakes scenarios, marking a significant stride toward their real-life applicability.

This paper presents a novel approach to SNN implementation, leveraging the strengths of FPGAs to address the challenges of image recognition in dynamic environments. The choice to focus on FPGA implementations for SNNs was driven by several considerations such as FPGAs offer a unique blend of flexibility and performance, allowing for rapid prototyping and iterative design that is crucial for the evolving field of SNNs. This work contributes to the field by not only validating the FPGA-based hardware design using a complex dataset but also by offering a comprehensive comparison with prior standard works. This research underscores the progression of neuromorphic computing \cite{Neuromorphic} towards practical, everyday applications, setting a new benchmark for future endeavors in the domain. In summary, the main contributions of this paper are:

\begin{itemize}
    \item Developing a vision-based TinyML framework optimized for tiny UAVs and UGVs.
    \item Adapting a widely-recognized $1^{st}$ Order LIF model for on-device processing in TinyML applications, tested with a challenging dataset.
    \item Architecting a robust hardware solution tailored for the efficient deployment of this SNN model on an FPGA platform.
\end{itemize}

\section{Related Works}
\label{sec:relwork}
Recent studies have focused on the integration of SNNs with ~FPGAs \cite{cite4,SNN_FPGA}. Inspired by biological neural networks, SNNs seek to mimic the information transmission and signal processing inspired by biological neurons \cite{cite1, cite2}. Neural dynamics play a major role in the physiologically realistic SNN's performance \cite{cite3}. FPGA implementations are becoming more popular because they can be used to create neural network systems with minimal power consumption, making use of FPGA's ability to create customized topologies for embedded and high-performance applications.

The development of an effective and adaptable FPGA architecture for the Izhikevich model-based real-time simulation of SNNs is a noteworthy achievement \cite{cite4}. With a 10 kHz sampling rate, this device can simulate up to 1,440 neurons in real-time, making it appropriate for small-to-medium-sized extracellular closed-loop research. Another aspect is the realization of more biologically plausible artificial neurons as proposed in \cite{cite5}. The implementation of online learning in neuromorphic chips is a challenging task that has been addressed to improve the functionality and flexibility of these systems. A bio-plausible online learning SNN designed for effective hardware implementation is shown in the suggested model. This advancement might completely change the way SNNs are utilized in various applications, including artificial intelligence and neuroprosthetics. It also represents a significant step towards building more robust and adaptable neuromorphic computing systems.

In recent research \cite{cite6,cite7,cite9,cite10,SNN}, neuromorphic systems have been developed primarily for pattern recognition tasks. The approach aims to create high-speed and large-scale systems. One notable implementation, \cite{cite6,cite7} has a hybrid updating algorithm, enhancing hardware design efficiency and performance. This system supports a network comprising 16,384 neurons and 16.8 million synapses while achieving a power consumption of 0.477 watts. Its effectiveness has been validated on a Xilinx FPGA, demonstrating a high accuracy of 97.06\% and a frame rate of 161 fps on the MNIST dataset. Another work is \cite{cite8}, which focuses on SNNs for real-time and energy-efficient applications, particularly in robotics and embedded systems. This model, consisting of an FCN with 800 neurons and 12,544 synapses, operates in real-time. It stands out for its event-driven architecture, which updates neurons based on spiking activity, leading to increased energy efficiency and reduced time resolution. Additionally, this design incorporates a simplified form of Spike-Time Dependent Plasticity (STDP) learning, balancing computational efficiency with robust learning capabilities.

While SNNs have been successfully implemented for various real-world applications, including sensory fusion, their deployment in real-time image recognition scenarios often poses unique challenges \cite{FPGA_signal1,FPGA_signal2}. These include not only the computational complexities inherent in processing dynamic visual data but also the need for tailored hardware optimization to accommodate the nuanced temporal dynamics of SNNs. Our current work aims to address these challenges by focusing on a more demanding collision avoidance dataset, which better reflects the complexities encountered in real-world environments than traditional datasets such as MNIST and their implementation on FPGA platforms. 
The aim is to validate the practicality of the hardware design using the LIF model in more demanding and realistic settings. This approach not only demonstrates the proof of concept for the hardware design but also facilitates a comprehensive comparison with previous standard works, marking a significant leap in the field's progression towards real-life applications.


\begin{figure}[t]
    \centering
    \scalebox{1}{
    \includegraphics[width=0.47\textwidth]{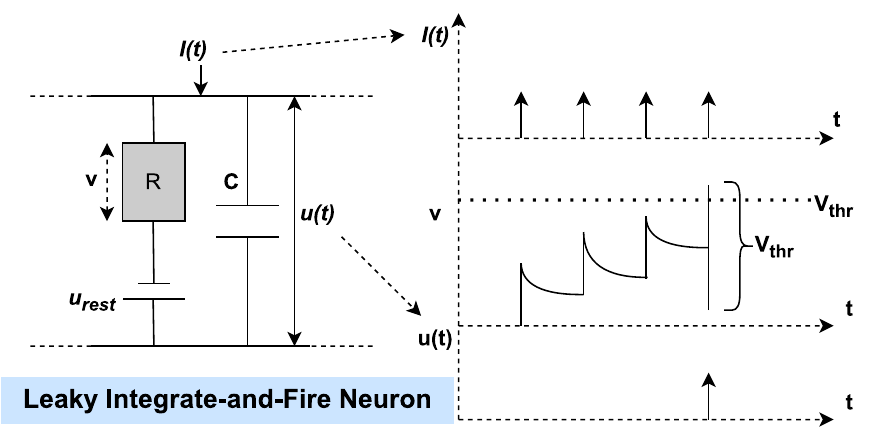}}
    \caption{RC Circuit Analogue of a 
    LIF neuron that emulates the behavior of a neuron. The membrane potential's response to a step input current highlights the voltage threshold for spike initiation and the subsequent reset.}
    \label{fig:RC_circuit_SNN}
    \vspace{-10pt}
\end{figure}

\section{Background and Input Coding}
\label{sec:Input_Encoding}


\subsection{SNN Neuron Model}
Various neuron models for SNN, like the Hodgkin-Huxley \cite{Hodgkin_Hulex}, Izhikevich \cite{Izhikevich}, Wilson \cite{Wilson}, Fitzhugh-Nagumo \cite{Fitzhugh-Nagumo}, Lapicque \cite{lapicque}, and Leaky Integrate-and-Fire \cite{LIF} models, offer different levels of complexity and biological fidelity. Among these, the Leaky Integrate-and-Fire (LIF) model is recognized for balancing computational simplicity with a reasonable degree of biological accuracy.

\textbf{Lapicque Model:} While the Lapicque model—often seen as the original form of the integrate-and-fire model—serves as a fundamental abstraction of neuronal firing by integrating incoming signals until a threshold is reached, it does not account for the leakage of charge through the neuron's membrane. The LIF model extends Lapicque's concept by introducing a leak term, which reflects the real neuron's property to lose some of its membrane potential over time due to its resistive properties. This addition provides a closer approximation to biological neurons, making the LIF model slightly more complex and bio-realistic. The membrane potential update equation for the Lapicque neuron model can be written as:

\begin{equation}
\label{equation: Lapicque}
    U[t+1] = U[t] + \frac{T}{C}I_{in}[t]
\end{equation}

where, $I_{in}$ - Input current, $U$ - Membrane potential, $U_{thr}$ - Membrane threshold, $T$ - duration of each time step, $R$- Reset mechanism and $\beta$ - Membrane potential decay rate.

If $U[t]>U_{thr}$, then a spike is emitted, and the potential is reset: $U[t+1]=0$. In contrast, the LIF model incorporates the membrane potential decay with a factor of beta ($\beta$), providing the following update equation:

\begin{equation}
\label{equation: LIF}
    U[t+1]=\beta U[t]+I_{in}[t+1]-R(U[t]+I_{in}[t+1])
\end{equation}
Upon reaching or surpassing the threshold the LIF model also emits a spike, but with the inclusion of a reset mechanism the membrane potential is reset to a baseline value: $U[t+1]=0$. Equation \ref{equation: LIF}
illustrate LIF's capability to simulate the gradual decay of membrane potential, which is characteristic of biological neurons and not represented in the simpler Lapicque model. The greater biological realism of the LIF model thus allows it to capture more complex dynamics of real neuronal behavior, particularly in the context of SNNs designed for tasks that may benefit from such nuanced representations.

\textbf{LIF Model:} The RC circuit depicted in Figure~\ref{fig:RC_circuit_SNN} serves as an electrical analog for the LIF neuron model. In this model, the neuron is represented by a resistor ($R$) and a capacitor ($C$) in parallel, analogous to the cell membrane's leak resistance and capacitance, respectively. The input current, $I(t)$, symbolizes the synaptic currents from presynaptic neurons. As the input current enters the circuit, the capacitor charges up, causing the membrane potential, modeled by the voltage across the capacitor $u(t)$, to increase. The resistor allows for leakage of charge, mimicking the imperfect insulation of a neuron's cell membrane, which causes the potential to decay over time back towards a resting potential ($u_{rest}$) if no further input is received.

When the potential $u(t)$ reaches a certain threshold $(V_{thr})$, the neuron fires, and this is represented by a spike in voltage. After firing, the potential is reset, typically to the resting potential, and the process can begin anew with further input. 
\begin{equation}
\label{equation: standard LIF}
    \tau_m \frac{du}{dt} = -[u(t)-u_{rest}]+RI(t)
\end{equation}

The dynamics of the membrane potential are governed by the differential equation \ref{equation: standard LIF}, which describes how $u(t)$ changes over time-based on the decay towards resting potential and the input current, with the membrane time constant $\tau_m=RC$ characterizing the rate of decay. This model simplifies the complex behavior of neurons by reducing action potentials to discrete events and does not account for the detailed shape of the spikes, focusing instead on their timing and occurrence to convey information.

\subsection{Input Coding}

In SNN coding technique is essential to convert static input data, such as images, into a dynamic, time-varying format that SNNs can process effectively. There are various coding techniques with rate coding being one of several methods. Aside from rate coding, time-to-first-spike (TTFS) coding is another method where the time at which the first spike occurs encodes the signal's intensity, emphasizing temporal precision. Delta modulation, on the other hand, encodes the change in input values over time, thereby capturing dynamic variations in the sensory signal. For our implementation rate coding was chosen for its simplicity and robustness, particularly when dealing with static datasets.

 One popular method for this transformation is rate coding \cite{Rate_coding} which is particularly suitable for static datasets, as it facilitates the representation of static input values as temporal spike patterns, which are the fundamental processing units in SNNs.

\begin{figure}[t]
    \centering
    \scalebox{1}{
    \includegraphics[width=0.47\textwidth]{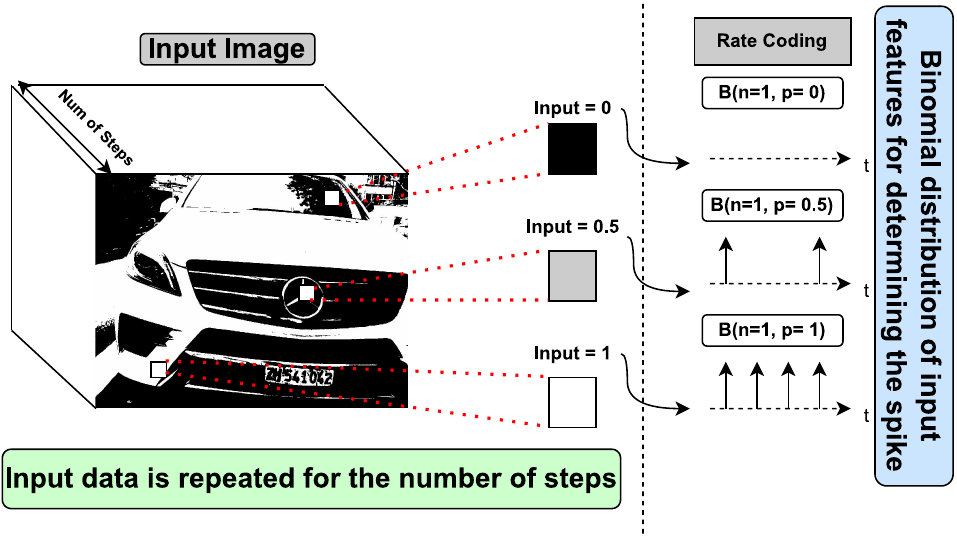}}
    \caption{ Visualization of Rate Coding for SNN. 
    An input image of a vehicle is converted into spike rates, with varying input intensities: absent (0), medium (0.5), and full (1). These intensities correlate with the density of a Bernoulli-distributed spike train over time, illustrating the transformation from pixel values to temporal spike patterns.} 
    \label{fig:Rate_coding}
    \vspace{-10pt}
\end{figure}

Rate coding operates on the principle that the information in a static input is translated into the frequency or rate of spikes over a given time interval. In essence, it is a way of mapping the intensity of an input signal 
to the frequency of spike events in a neuron which is shown in Figure~\ref{fig:Rate_coding}. The key aspects of rate coding include:
\begin{enumerate}
\item Pixel Intensity to Spike Rate: Each pixel's intensity in an image is correlated with the spiking rate of a neuron. In a grayscale image, pixel intensities range from black (low intensity) to white (high intensity).
\item Black Pixel Representation: 
Indicating lower intensity, are less likely to generate spikes, 
corresponding to dead neurons that do not fire. This suggests that darker areas of an image may correspond to less frequent neuron firing.
\item White Pixel Representation: 
White or lighter pixels, which represent higher intensities, are associated with a higher probability of generating spikes. This implies that brighter areas in an image will result in more active neuron firing.
\end{enumerate}

The differential spiking activity based on pixel intensity is crucial for feature differentiation within the visual input. However, to implement rate coding in a practical scenario we follow the following steps: 
(1) Normalization: Pixel values are first normalized, typically to a range between 0 and 1. This normalization is crucial for consistent spike generation across different images and lighting conditions.
(2) Spike Generation Probability: The normalized pixel value determines the probability of spike generation at each time step within the coding window. For example, a pixel value of 0.8 might mean there is an 80\% chance of a neuron firing at each time step.
(3) Temporal Aspect: The coding window is divided into discrete time steps. At each time step, a spike is generated with a probability corresponding to the pixel's normalized intensity.
(4) Spike Train Formation: For each pixel, this process results in a spike train over the coding window, representing the temporal pattern of neural activity corresponding to that pixel's intensity.
\section{Proposed Approach}
\label{sec:propapp}

\begin{figure}[t]
    \centering
    \scalebox{1}{
    \includegraphics[width=0.47\textwidth]{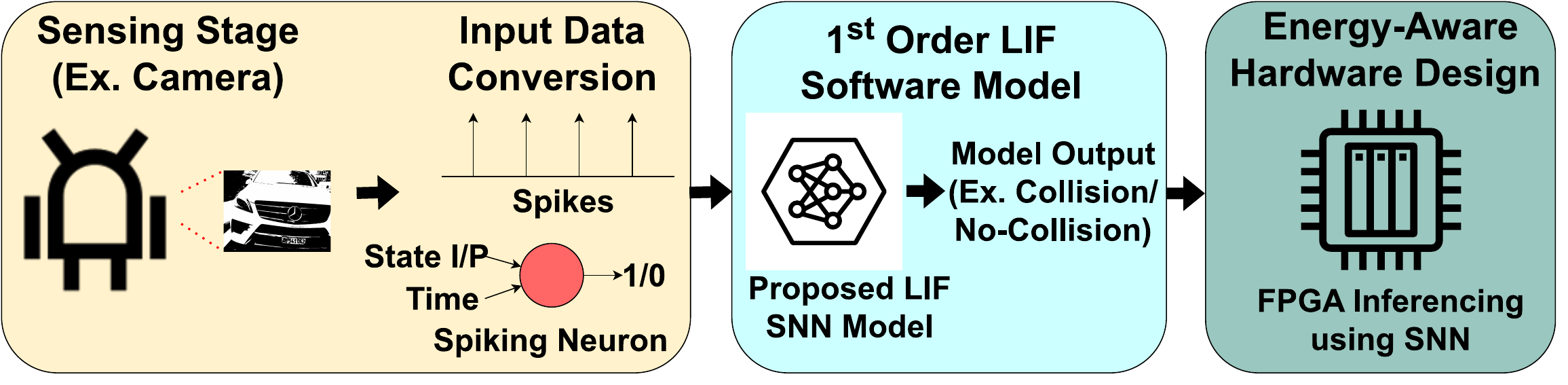}}
    \caption{This image depicts a three-stage \Mozhgan{energy-aware framework for TinyML systems: }
    A camera captures the scene (Sensing), visual data is translated into spikes by a neuron model (Input Data Conversion), an SNN software model predicts collision events (Software Model), and the SNN model is implemented on an FPGA 
    (FPGA Deployment).}
    \label{fig:System_SNN}
    \vspace{-10pt}
\end{figure}


\Mozhgan{The diagram in Figure~\ref{fig:System_SNN} illustrates an energy-aware SNN framework as a TinyML approach for resource-constrained systems. This framework can be applied to various vision-based applications such as autonomous navigation for detecting potential collisions using an enhanced SNN model. The proposed framework has three stages including: sensing stage and input data conversion, $1^{st}$~LIF software model, and energy-aware hardware design.
}


\subsection{Sensing Stage and Input Data Conversion} A camera serves as the sensory input device, capturing 
imagery of the surrounding environment. 
\Mozhgan{This image data includes useful information based on the application, for instance potential obstacles or conditions that could lead to a collision.}


The raw camera data undergoes a transformation where the pixel intensities of the images are converted into spike rates. This conversion is illustrated by spiking neurons, which convert the state input into a time-dependent spike output. The spiking neurons are shown to emit spikes over time, indicating dynamic information processing based on the input data.



\subsection{$1^{st}$ Order Leaky Integrate-and-Fire (LIF) Software Model}
\subsubsection{Spiking Neural Network Model}
\begin{figure}[t]
    \centering
    \scalebox{1}{
    \includegraphics[width=\columnwidth]{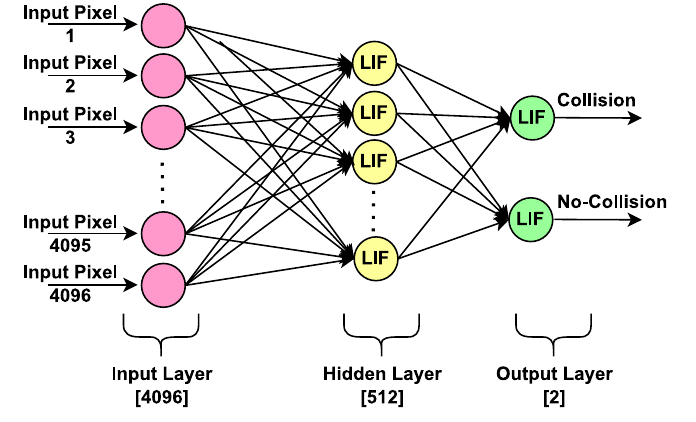}}
    \caption{The proposed architecture of 
    SNN software model for \Mozhgan{TinyML applications such as} collision detection \Mozhgan{onboard}, featuring an input layer with 4096 neurons corresponding to image pixels, a hidden layer with 512 
    LIF neurons, and an output layer indicating collision or no-collision.}
    \label{fig:FCN_SNN}
    \vspace{-10pt}
\end{figure}

This section describes the implementation of an enhanced SNN model, \Mozhgan{designed for TinyML applications such as autonomous navigation and} collision avoidance in tiny UAVs, UGVs and other \Mozhgan{resource-constrained devices in terms of computation capacity, available memory and power consumption.}
As shown in Figure~\ref{fig:FCN_SNN} the SNN model comprises three main layers:
(1) Input Layer: linear transformation that flattens the input images into a vector of size equivalent to the image's pixel count (64$\times $64). 
(2) Hidden Layer: Leaky Integrate-and-Fire neurons, with learnable parameter such as, threshold and $\beta$. Additionally, dropout is applied for regularization.
(3) Output Layer: linear transformation followed by leaky neurons, producing a final output corresponding to the two classes (collision, non-collision).

The model operates over 25 time steps, integrating the temporal aspect of spiking neurons. This allows the network to process inputs as a series of spikes over time, which is a fundamental characteristic of SNNs. The model is trained using the Adam optimizer \cite{adam} with a learning rate of $5e^{-4}$. Cross-entropy loss is computed across all time steps, summing up to form the total loss for each training instance. The training process involves forward passes through the network over multiple epochs, with loss calculation and backpropagation at each step. The network's performance is evaluated in terms of accuracy on both the training and testing sets. 


\subsubsection{SNN with Refractory Period}
\Mozhgan{To further improve the proposed SNN model for TinyML systems, a refractory period is implemented on the proposed $1^{st}$ LIF SNN software model.}
The refractory period is a critical feature in biological neurons where, after firing, a neuron becomes less sensitive or unresponsive for a short duration. This implementation aims to mimic this biological behavior, leading to more robust and biologically plausible SNN behavior.

\textbf{Refractory Mechanism.} The refractory period is implemented by suppressing the firing of a neuron for a specified number of time steps after it spikes. This is achieved by maintaining a record of the last spike time for each neuron and preventing subsequent spikes. 

\textbf{Parameterization.} The refractory period is set to 5-time steps, though this can be adjusted based on experimental requirements.

\textbf{Network Layers.} The network architecture is similar 
with the addition of refractory logic in both the hidden and output layers.

The implementation of a refractory period in SNNs represents a step towards more biologically inspired ANN. 
This model stands as a testament to the adaptability and versatility of SNNs in mimicking the intricate functionalities of the brain, with potential applications in areas requiring efficient and dynamic information processing.

\begin{figure}[t]
    \centering
    \scalebox{1}{
    \includegraphics[width=\columnwidth]{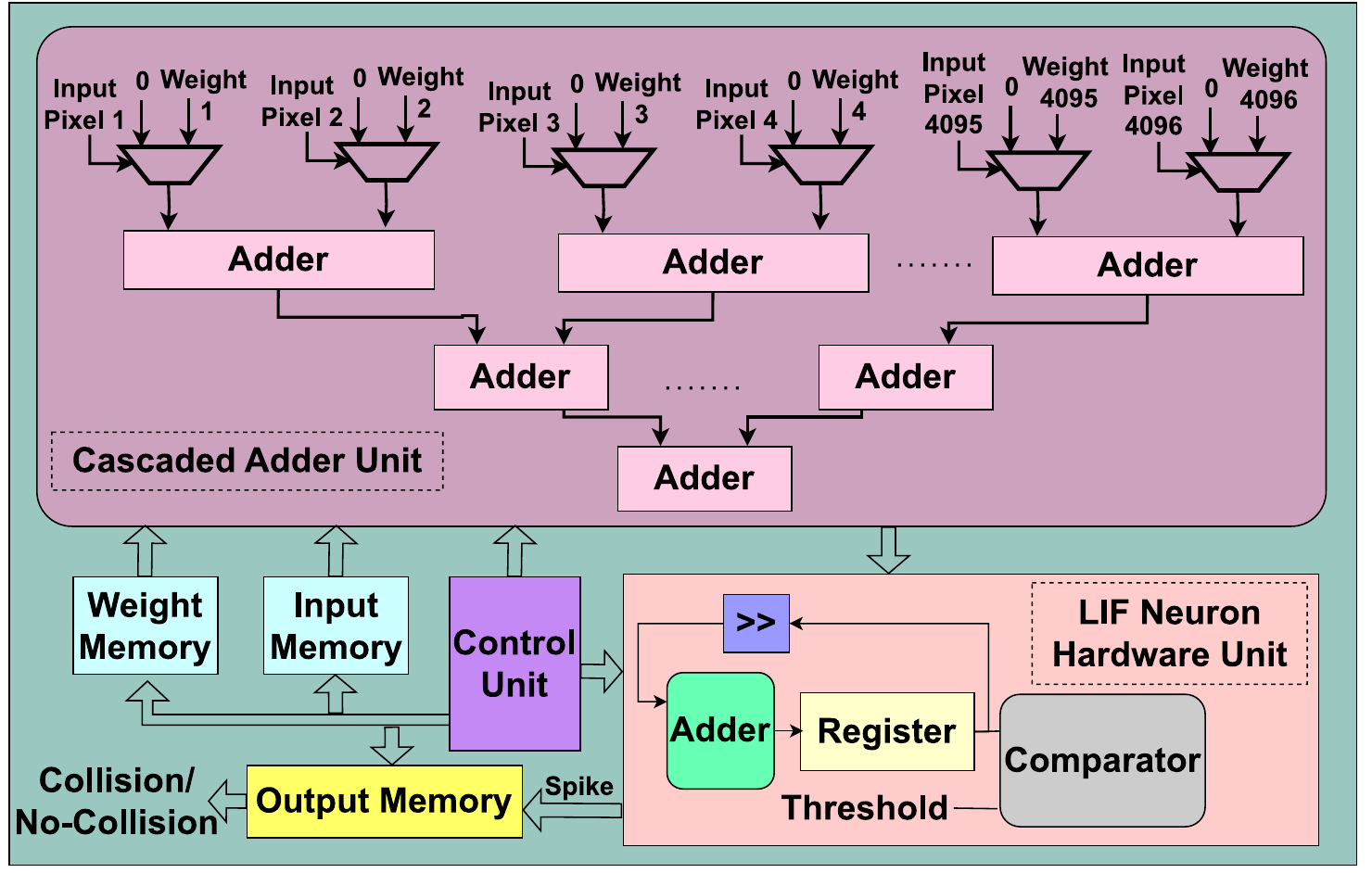}}
    \caption{Figure illustrating the proposed flow of hardware implementation for a Spiking Neural Network, integrating cascaded adders for input processing and a Leaky Integrate-and-Fire (LIF) neuron model with a control unit, input and weight memory, and output classification.}
    \label{fig:FPGA_dep}
    \vspace{-10pt}
\end{figure}

\begin{figure*}[t]
    \centering
    \scalebox{1}{
    \includegraphics[width=\textwidth]{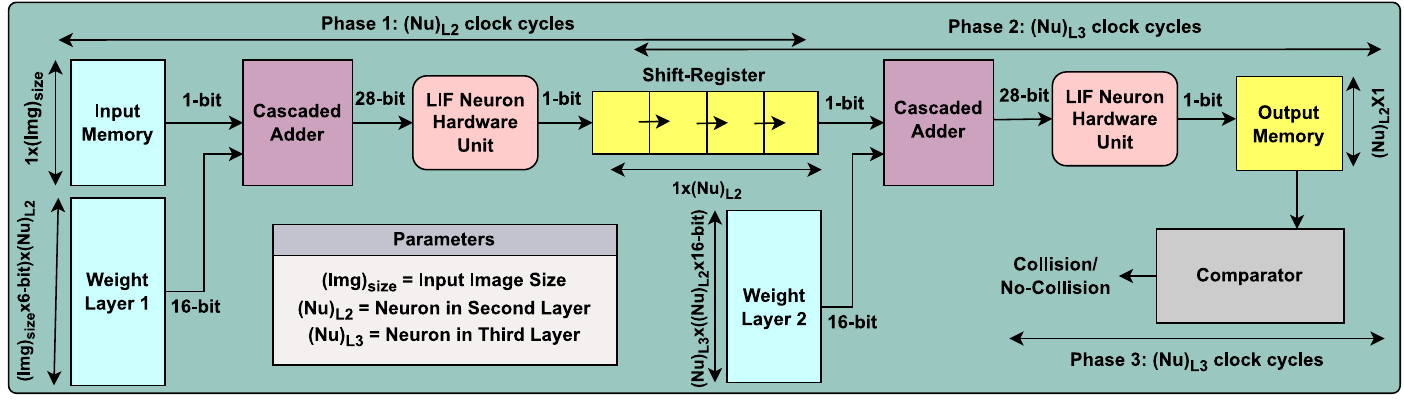}}
    \caption{Schematic representation of the hardware implementation phases for the proposed SNN
    , showing the data flow through weight layers, cascaded adders, LIF neurons, and a comparator for class prediction, with the process segmented into specific clock cycle phases.}
    \label{fig:SNN_flow}
    \vspace{-10pt}
\end{figure*}



\subsection{Energy-Aware Hardware Design} 


Implementing the SNN architecture on an FPGA harnesses the unique characteristics of FPGAs for efficient neural computation. In deploying an SNN  \Mozhgan{on resource-constraint systems} 
a strategic shift is made towards binary inputs and adder-based operations as shown in Figure~\ref{fig:FPGA_dep}. Binary input encoding yields significant reductions in memory bandwidth and storage compared to a traditional Fully Connected 
Network~(FCN) or Convolutional Neural Network~(CNN) that typically uses 16-bit representations for weights and inputs. The SNN takes advantage of the binary format by substituting the conventional Multiply-Accumulate~(MAC) operations with a cascaded adder network, which outputs a more resource-efficient 28-bit intermediate result, rather than the 32-bit output as seen in FCN and CNN. This streamlined data flow is further optimized by the incorporation of LIF neurons, which process the output from the cascaded adder and convert it into binary spikes. By implementing output memory as a shift register, the design conserves resources more effectively than the larger memory arrays required for the multi-bit outputs of an FCN or CNN. The culmination of these features, an adder-centric approach, and binary input-output processing, not only reduces the number of operations but also minimizes latency and power usage, thereby enhancing the overall efficiency of the SNN model on FPGA platforms \Mozhgan{for TinyML.}

In 
\Mozhgan{the proposed $1^{st}$ LIF} SNN model, weights, biases, and neuron states are represented using a 16-bit signed fixed-point format, Q1.15. This format allows for a more precise range. 
The increased bit depth in Q1.15 enables finer resolution in representing fractional values, which is crucial for capturing subtle variations in 
parameters.
With the adoption of the Q1.15 format, all neural network calculations are confined within the -1~to~+1~range. This constraint is vital to maintain data representation consistency and to prevent overflow or underflow errors in computations. The wider range of Q1.15 allows for a more detailed and accurate representation of the 
parameters, enhancing the overall precision of the computations without compromising the operational limits.
For the computation required for a single neuron, the architecture is divided into two parts, namely Cascaded Adder and LIF Neuron Hardware Unit.

\textbf{Cascaded Adder Unit.} A cascaded adder is used for FPGA deployment of SNN to efficiently sum binary inputs, thereby optimizing resource usage and power consumption.
Given that the input pixel values are binary (0 or 1), the standard matrix multiplication operation common in neural networks is simplified. A cascaded adder, or an adder tree, replaces the need for multipliers, which are traditionally more resource-intensive on silicon. This substitution significantly reduces the hardware complexity and power consumption.
The addition of bias values to the neurons is performed after the cascaded adder computation, maintaining the integrity of the neural network's functionality.


\textbf{LIF Neuron Hardware Unit.} The LIF neuron hardware unit 
is an integral part of the FPGA-based SNN architecture, designed to emulate the dynamic behavior of biological neurons in a digital framework. Its primary functionality revolves around processing input signals, accumulating them over time, and generating an output spike under certain conditions, akin to a neuron firing in response to stimuli as given in equation \ref{LIF equation}.
\begin{equation}
U[t+1]={\beta}U[t] + I[t+1] - U_{rest}
\label{LIF equation}
\end{equation}

This unit simulates the neuron's response to stimuli, where input signals are accumulated until a threshold is reached, leading to a spike. This process mirrors the all-or-nothing firing mechanism of biological neurons. The use of a threshold in determining the firing of the neuron is significant. It ensures that the NHU only generates a spike when the accumulated input is sufficiently strong, thereby mimicking the selective response of neurons in the brain. The reset of a spike is a crucial aspect of the NHU, as it models the refractory period in biological neurons. This mechanism prevents immediate re-firing, ensuring a realistic emulation of neuronal behavior.



\section{Experimental Results}
\label{sec:exp}


\begin{table}[t]
\centering
\caption{SNN Performance by Image Size and Neuron Model}
\vspace{-10pt}
\label{tab:performance}
\resizebox{\columnwidth}{!}{%
\begin{tabular}{|c|c|c|c|c|}
\hline
\textbf{Image Size} & \multicolumn{2}{c|}{\textbf{LIF Model}} & \multicolumn{2}{c|}{\textbf{Lapicque Model}} \\ \cline{2-5} 
                                     & \textbf{Train Acc.} & \textbf{Test Acc.} & \textbf{Train Acc.} & \textbf{Test Acc.} \\ \hline
32$\times $32                                & 93\%                        & 79\%                      & 93\%                       & 84\%                     \\ \hline
\textbf{64$\times $64}                                & \textbf{92\%}                        & \textbf{85\%}                      & 95\%                         & 81\%                       \\ \hline
128$\times $128                              & 88\%                        & 78\%                       & 92\%                      & 80\%                      \\ \hline
\end{tabular}}
\vspace{-10pt}
\end{table}

\subsection{Experimental Setup}

The SNN model is built using the snntorch library, a specialized Python package tailored for simulating spiking neural networks. It integrates smoothly with PyTorch's ecosystem for GPU-based acceleration of training and testing the model. The input images are taken from a dataset \cite{dronet} consisting of approximately 32K annotated images, each labeled with collision or no-collision to train the network during the supervised learning process. Preprocessing steps include resizing the images to a uniform dimension of 64$\times$64 pixels, converting them to grayscale to reduce computational load, and normalizing the pixel values. These preprocessed images are then fed into the model using PyTorch's DataLoader, which facilitates efficient batch processing and data shuffling.

For the implementation of the LIF neuron model on FPGA, we code it in Verilog HDL to simulate the SNN which is compiled using the Xilinx ISE to ensure optimization for hardware application. The resulting SNN is deployed on the Xilinx Artix-7 FPGA 
board.

\subsection{Software Results}
In this section, we analyze the performance of two neuron models, LIF and Lapicque, across three different image sizes: 32$\times $32, 64$\times $64, and 128$\times $128 pixels (Table \ref{tab:performance}). The LIF model demonstrated a training accuracy that slightly declined with increasing image size, starting at 93\% for 32$\times $32 images and decreasing to 88\% for 128x128 images. Testing accuracy for the LIF model showed a similar trend, with the highest accuracy of 85\% for 64$\times $64 image size. The observed decline in the LIF model's performance with increasing image size because larger images introduce more variability and require more complex feature detectors than the current architecture of the LIF model moreover, as the image size grows, hyperparameters might not scale well, leading to less efficient learning and lower accuracy.

The Lapicque model, conversely, displayed more consistent performance across the tested resolutions, achieving up to 95\% training accuracy and maintaining above 80\% in testing accuracy for all sizes, peaking at 84\% for 32$\times $32 images. Despite this, the LIF model is often preferred for hardware implementations where mimicking biological neuron behavior is critical, such as in real-time collision avoidance systems. In these scenarios, the event-driven and energy-efficient nature of SNNs, akin to biological brains, is especially advantageous. It ensures computational activities are executed in response to distinct stimuli, offering significant energy savings—essential for battery-operated or energy-limited platforms like UAVs and UGVs.


\begin{table}[t]
\caption{Comparison of SNN and BCNN on FPGA}
\vspace{-10pt}
\centering
\resizebox{\columnwidth}{!}{%
\begin{tabular}{|c|c|c|}
\hline
\textbf{Metric} & \textbf{SNN (This work)} & \textbf{BCNN (Baseline) \cite{BCNN_compare}} \\
\hline
FPGA Platform & Artix-7 & Zynq-7000 \\
\hline
Power (mW) & \textbf{495} & 2300 \\
\hline
Performance (GOPS) & \textbf{541} & 329 \\
\hline
Frequency (MHz) & \textbf{67} & 143 \\
\hline
Energy Efficiency (GOPS/W) & \textbf{1093} & 143 \\
\hline
\end{tabular}}
\label{tab:comparison}
\vspace{-10pt}
\end{table}

\begin{table*}[t]
\caption{Comparison of Proposed Neuron Model With Previous Works}
\vspace{-10pt}
\centering
\begin{tabular}{|c|c|c|c|c|c|}
\hline
\textbf{Neuron Model} & \textbf{Slice Registers} & \textbf{Slice LUTs} & \textbf{Frequency (MHz)} & \textbf{Power (mW)} & \textbf{Device}\\
\hline
Spike Response Model \cite{Ref1}& 202 & 378& 100&-&Spartan3 XC3S400\\
\hline
Wilson \cite{Ref6}& 365 & 611& 98 &-&Virtex-6 ML605\\
\hline
Izhikevich \cite{Ref5}& 31 & 771 & -& 1043&Virtex-5\\
\hline
Simplified LIF \cite{Ref2}& 29 & 70& 100&-&Virtex 6\\
\hline
LIF \cite{Ref3}& 297 & 196 & 100&0.48&Virtex-7 VX690T\\
\hline
LIF \cite{SNN}& 16 & 27 & 50&95&Artix-7\\
\hline
\textbf{$1^{st}$ Order LIF (This Work)} & \textbf{96} & \textbf{25} & \textbf{100} & \textbf{85}&Artix-7\\
\hline
\end{tabular}
\label{tab:compare}
\end{table*}

\begin{table*}[t]
\caption{Comparision of Proposed SNN with Previous Works}
\vspace{-10pt}
  \centering
  \begin{tabular}{|c|c|c|c|c|c|}
    \hline
    \textbf{Reference} & \textbf{Slice Register} & \textbf{Slice LUT} & \textbf{Frequency (MHz)} & \textbf{Architecture} & \textbf{Device} \\
    \hline
    \cite{Ref4} & 1023 & 11339 & 189 & 25-5-1 & Virtex-6 ML605 \\
    \hline
    \cite{SNNcompare2} & 33 & 19 & 717 & 25-10 (interconnection) & Terasic DE1-SoC \\
    \hline
    \cite{SNN} & 119 & 587 & 25 &25-5-2 & Artix-7 \\
    \hline
    \textbf{This Work} & \textbf{1780} & \textbf{6521} & \textbf{67} &4096-512-2 & Artix-7 \\
    \hline
  \end{tabular}
  \label{tab:SNN_compare}
\end{table*}

Through our analysis, we aim to demonstrate that the LIF model, with its closer alignment to biological neuron dynamics, offers distinct benefits for applications like collision avoidance. This is due to its real-time responsiveness and ability to adaptively process sensory information, ensuring efficient and safe navigation in complex environments.


\subsection{FPGA Deployment Results}

From the Table~\ref{tab:comparison}, it is evident that the SNN model is significantly more power-efficient than the BCNN. The SNN operates at a power consumption of 495~mW, which is considerably lower than the BCNN. 
Despite the lower power usage, the SNN achieves a higher performance in terms of GOPS, with a value of 541 compared to the BCNN's 329~GOPS. This suggests that the SNN is not only using less power but is also delivering more computational operations per second. Furthermore, the energy efficiency metric, measured in GOPS/W, highlights a contrast between the two models. The SNN demonstrates an energy efficiency of 1093~GOPS/W, which is 
86\% higher than the BCNN. 
This indicates that the SNN can perform more operations for each watt of power consumed, making it an extremely energy-efficient architecture compared to the BCNN.

The detailed results from the hardware implementation of various neuron models on different FPGA devices are summarized in Table \ref{tab:compare}, which facilitates a direct comparison of the proposed $1^{st}$ Order LIF neuron model with prior works. This comparison reveals the proposed LIF neuron model's strategic balance between resource utilization, operational frequency, and power efficiency. Particularly, when compared to the Spike Response Model \cite{Ref1} and the Izhikevich neuron model \cite{Ref5}, the proposed LIF model demonstrates moderate use of resources, reflected by the number of slice registers and Look-Up Tables~(LUTs) required. These models, while operating at similar or lower frequencies, show a higher consumption of FPGA resources, which could imply greater costs or larger physical footprints in deployment.

In contrast to the LIF model \cite{SNN}, which indeed shows less resource utilization, the proposed $1^{st}$ Order LIF model operates at a doubled frequency of 100~MHz and exhibits lower power consumption at 85~mW. This enhancement in frequency suggests an improved capability to process information rapidly, which can be critical for time-sensitive applications. Moreover, the reduced power usage of the proposed model extends the potential for deployment in power-sensitive contexts, such as battery-operated embedded systems or portable devices requiring efficient neural network computation.

Table~\ref{tab:SNN_compare} provides a comparative analysis of the proposed SNN implemented against various other SNN implementations from previous studies. It showcases the proposed SNN's resource usage in terms of slice registers and LUTs, the operating frequency of the FPGA, and the specific FPGA device used for each implementation. The operating frequency of the proposed SNN is 67~MHz, which is moderately placed among the referenced works. It is notably lower than the highest frequency presented, but this is a strategic choice to balance power consumption and processing speed, which is often a critical consideration in embedded and real-time applications. The comparison shows that while the proposed SNN requires more resources than some of the simpler models, it still operates effectively within the constraints of the Artix-7 device. Due to the variations in SNN architectures, FPGA platforms, and versions of synthesis tools, the results presented in Table~\ref{tab:SNN_compare} should be considered as estimations.

\section{Conclusion}
\label{sec:con}


\Mozhgan{In this research 
, we introduced a new framework called $1^{st}$ order LIF SNN, which aims to efficiently deploy vision-based ML algorithms on TinyML systems.
The proposed approach involves training an innovative obstacle detection model using SNN architectures, with a focus on LIF neurons. Additionally, we devised a highly optimized hardware setup specifically tailored for this approach and successfully implemented it using a Xilinx Artix-7 FPGA. To assess the effectiveness of our approach, we conducted a comparative analysis between the accuracy of the proposed SNN model and different image input sizes for two types of models, namely LIF and Lapicque. Furthermore, to measure power consumption and resource utilization, we developed the hardware and implemented the one that demonstrated superior accuracy on an FPGA board. The experimental results indicate that the $1^{st}$ order LIF approach is 
achieving up to 86\% greater energy efficiency compared to a Binarized CNN. Moreover, our approach achieved an impressive accuracy rate of up to 92\% with $64\times64$ image size for LIF model.}



\bibliographystyle{ACM-Reference-Format}
\bibliography{acmart}
\end{document}